\documentstyle[amssymb,aps]{revtex}
\draft

\begin{document}

\tighten

\twocolumn[\hsize\textwidth\columnwidth\hsize\csname@twocolumnfalse\endcsname

\title{Fragmented Condensate Ground State of Trapped Weakly Interacting Bosons \\
in Two Dimensions}
\author{Xia-Ji Liu$^1$, Hui Hu$^1$, Lee Chang$^2$, Weiping Zhang$^1$, Shi-Qun Li$^{1,3}$,
and Yu-Zhu Wang$^4$}
\address{$^1$Department of Physics,Tsinghua University, Beijing 100084, China\\
$^2$Center for Advanced Study, Tsinghua University, Beijing 100084, China\\
$^3$Center for Atomic and Molecular Nanosciences, Tsinghua University, Beijing
100084, China\\
$^4$Shanghai Institute of Optics and Fine Mechanics, Chinese Academy of
Sciences, Shanghai 201800, China}
\maketitle

\begin{abstract}
The ground state and its structure for a rotating, harmonically trapped $N$-Boson
system with a weak repulsive contact interaction are studied as
the angular momentum $L$ increases up to $3N$. We show that the ground state
is generally a fragmented condensate due to angular momentum
conservation. In response to an (arbitrarily weak) asymmetric perturbation
of the trap, however, the fragmented ground state can be transformed into a
single condensate state. We manifest this intrinsic instability by
calculating the conditional probability distributions, which show patterns
analogous to the boson density distributions predicted by mean-field theory.
\end{abstract}

]

Following the experimental realization of a dilute atomic Bose-Einstein
condensate (BEC), the formation and properties of vortices in an atomic BEC
have caused considerable interest both experimentally \cite{expt} and
theoretically \cite{TF,attr,mott,bertsch,com,mft,mft-vs-ex,analytic} in the
past few years. Although the recent demonstrations of vortex states by
several different groups are in the Thomas-Fermi limit of strong interatomic
interaction, a great deal of attention have also been attached to the
nonvanishing angular momentum states of weakly interacting $N$-Boson systems
in harmonic traps. In Ref. \cite{attr}, Wilkin {\it et al}. considered the
case of an attractive interaction and showed that the ground state is {\em %
uncondensed} and is an example of the ''{\em fragmented}'' condensate
discussed by Nozi\`{e}re and Saint James \cite{nozieres}. Mottelson \cite
{mott}, Bertsch and Papenbrock \cite{bertsch} considered the lowest energy
quantum states of a repulsively interacting Bose gas when $L\leqslant N$.
Wilkin {\it et al}. have employed a composite boson/fermion picture to
describe configurations beyond the one-vortex state \cite{com}. A more
tractable mean-field calculation performed by Butts and Rokhsar revealed a
succession of transitions between stable vortex patterns of differing
symmetries in the high angular momentum regime \cite{mft}. The connection
between the mean-field theory (MFT) and exact diagonalization scheme has
been studied by Jackson {\it et al}. for a special case of $L=2N$ \cite
{mft-vs-ex}. Finally, some analytical results have also been reported for
the lowest energy states \cite{analytic}.

In this Letter we address the question of whether the ground state of a
weakly interacting $N$-Boson system with a given angular momentum is what
one would normally expect, i.e., a state with a single coherent Bose
condensate, in which a mean-field approximation is valid. We propose that
this is not the case and the ground state is generally a {\em fragmented}
condensate in the presence of the weakly repulsive interatomic interaction
except $L/N=0$ or $1$ in the thermodynamic limit. By evaluating the
macroscopic eigenvalues of the single-particle (SP) density matrix, we
determine the degree of condensation. The origin of fragmentation turns out
to be a requirement of the conservation of angular momentum. As a result, by
turning on an (arbitrarily weak) asymmetric perturbation of the trap, the
fragmented ground state can be easily deformed to a single condensate state 
\cite{rokhsar98}. This intrinsic instability can be manifested by the
conditional probability distributions (CPDs) calculated for the ground
state, which show patterns analogous to boson density distributions
predicted by MFT. Note that the weakly interacting $N$-Boson system
considered here are quite similar to the spin-1 Bose gas studied by Ho and
Yip \cite{ho}, in which the fragmentation originates from the spin
conservation.

We start from the model Hamiltonian describing $N$ bosons in a
two-dimensional harmonic trap interacting via a weak contact interaction.
The SP spectrum is usually expressed in terms of the angular momentum
quantum number $m$ and the radial quantum number $n_r$, by $E_{n_r,m}=\left(
2n_r+\left| m\right| +1\right) \hbar \omega .$ In the ground state of the
system all the bosons are in states with $n_r=0$, and with $m$ being zero or
having the same sign as the total angular momentum. In the second quantized
form, the Hamiltonian reads 
\begin{eqnarray}
{\cal H} &=&{\cal H}_0+{\cal V},  \nonumber \\
{\cal H}_0 &=&\hbar \omega \sum\limits_j(j+1)\hat{a}_j^{+}\hat{a}_j, 
\nonumber \\
{\cal V} &=&\frac 12g\sum\limits_{i,j,k,l}V_{ijkl}\hat{a}_i^{+}\hat{a}_j^{+}%
\hat{a}_k\hat{a}_l,
\end{eqnarray}
where ${\cal H}_0$ is the SP oscillator Hamiltonian and ${\cal V}$ is the
two-body interaction between bosons. In the perturbative regime of weak
interactions, $Ng\ll \hbar \omega $. The operator $\hat{a}_j$ and $\hat{a}%
_j^{+}$ annihilate and create one boson in the SP oscillator state $\left|
j\right\rangle $ with energy $(j+1)\hbar \omega $ and angular momentum $%
j\hbar $, respectively, and obey the bosonic commutation rules. The contact
interaction elements are given by $V_{ijkl}=\delta
_{i+j,k+l}2^{-(i+j)}(i+j)!/(i!j!k!l!)^{1/2}$ \cite{analytic}, and most of
them are actually vanishing. For a given total angular momentum $L$ and
number of bosons $N$, we consider the Fock space spanned by states $\left|
\alpha \right\rangle =\left| n_0,n_1,...,n_k\right\rangle $ with $%
\sum\limits_jn_j=N$ and $\sum\limits_jjn_j=L$. Here $n_j$ denotes the
occupation of the $j$th SP oscillator state $\left| j\right\rangle .$ There
is a huge degeneracy corresponding to many different ways of distributing $L$
quanta of angular momentum among $N$ atoms. Here we restrict ourselves in a 
{\em truncated} Fock space of $0\leqslant j\leqslant j_{\max }=6$ \cite
{mft-vs-ex,jmaxnote}. To obtain the energy spectra and the corresponding
eigenstates, we set up the matrix elements in the Fock space basis, and
subsequently diagonalize the matrix by using the Davidson algorithm \cite
{davidson}.

{\it A fragmented ground state.}---First consider the SP density matrix in
the form of 
\begin{equation}
\rho ({\bf r},{\bf r}^{\prime })=\sum\nolimits_{ij}\psi _i^{*}({\bf r})\rho
_{ij}\psi _j({\bf r}^{\prime })
\end{equation}
with $\psi _m({\bf r})=\left\langle {\bf r}\mid m\right\rangle $. In Ref. 
\cite{yang}, Yang showed that the appearance of condensation is associated
with the single macroscopic eigenvalue (i.e. of order $N$) of the density
matrix $\rho ({\bf r},{\bf r}^{\prime })$ with the ''condensate wave
function'' being the associated eigenvector, while the case of more than one
macroscopic eigenvalue has been referred to as a ''fragmented'' condensate 
\cite{nozieres}. The most important difference between the single and
fragmented condensate is the lack of phase coherence of the latter. To find
the eigenvalues of the SP density matrix, we write \cite{yang} 
\begin{equation}
\rho _{ij}={\rm Sp\quad }\hat{a}_i\rho \hat{a}_j^{+},
\end{equation}
where the trace runs over all the $N-1$ boson states, and the density matrix 
$\rho =\left| \Psi _{GS}\right\rangle \left\langle \Psi _{GS}\right| $. It
is readily seen that the eigenvalues are nothing but the occupation numbers
of the SP oscillator state due to the conservation of the total angular
momentum, namely $\rho _{ij}=\delta _{ij}n_j$. It is difficult to give a
explicit expression for the occupation numbers $n_j$. In the case of $L=N$,
Wilkin {\it et al}. find that in the limit of $N\rightarrow \infty ,$ to the
order $O(1/N)$ \cite{attr}, 
\begin{equation}
n_0=1,\quad n_1=N-2,\quad \text{and}\quad n_2=1.
\end{equation}
They therefore conclude that the $N$-Boson system is fully condensed into
the one-vortex state in the thermodynamic limit. More detailed information
can be obtained from the exact diagonalization calculations \cite{bertsch}.
In figure 1, we show the $L$ dependence of the occupation numbers $n_j$ and
their fluctuations $\Delta n_j=(<\hat{n}_j^2>-<\hat{n}_j>^2)^{1/2}$ for $%
j=0,1,2,3,4$ for a system of $N=40$ bosons. When $L\leqslant N$ the
occupation numbers evolve rather smoothly as the angular momentum increases 
\cite{bertsch}, while for $L>N$ there are many kinks in the curves,
reflecting the complexity of the ground states. The most prominent feature
in the figure is that for a high angular momentum there are generally at
least two {\em significant} occupation numbers. For instance, at $L=70$, the
system has two large occupation numbers: $n_j\approx 9$ and $23$ for $j=0$
and $2$, respectively. Evidently the case gives a ''fragmented'' condensate.
Although the present calculation is performed in the case of $N=40$, the
conclusion that a fragmented condensate ground state exists universally
applies to the trapped, weakly interacting and rotating $N$-Boson systems
with an arbitrary $N$ including the thermodynamic limit $N\rightarrow \infty 
$ \cite{mft}.

To examine the validity of the above statement, we investigate the $N$%
-dependence of the number of significantly occupied SP states by computing
the inverse participation ratio \cite{IPR,bertsch}: 
\begin{equation}
I_C=\sum\nolimits_j(n_j/N)^2.
\end{equation}
The $I_C$ is the first nontrivial moment of the distribution of occupation
numbers among the different SP states (note that $\sum\nolimits_jn_j/N=1$ by
normalization). Its inverse $1/I_C$ qualitatively measures the number of
significantly occupied SP states. For example, $I_C$ would be unity for a
system that only had a single macroscopic occupied states; the maximum value
of $N$ is reached in the opposite extreme, when all the SP oscillator states
of the system are equally occupied. Figure 2 shows a plot of $1/I_C$ as a
function of angular momentum $L$ for a system of $N=30,40,50$ and $60$
bosons. It is easy to see that in the regime of $L/N<1.6$ the value of $%
1/I_C $ varies smoothly as $L/N$ increases and shows little dependence on $N$%
. In particular, the variation of the peak height at $L/N\approx 1.6$ is
less than $3\%$ as $N$ increases from $30$ to $60$ (not shown in the
figure). For $L/N>1.6$, some irregular small oscillations appear in the
curves. However, the overall profile of $1/I_C$ is still nearly independent
of $N$. These small oscillations are purely due to the finite $N$ effect 
\cite{osc} and decay gradually with increasing $N$. One may expect them to
vanish in the limit of $N\rightarrow \infty $. Therefore, we conclude that $%
1/I_C$ can be further used to {\em qualitatively} measure the number of {\em %
macroscopically} occupied SP states in the thermodynamic limit, or in other
words, to determine whether the ground state is fragmented or not.

As shown in figure 2, there are two global minima ($\approx 1$) at $L/N=0$
and $L/N=1$, which can be well interpreted as a signature of single
condensates. For other values of $L/N$ (especially in the high angular
momentum), however, $1/I_C$ is generally larger than $2$. This clearly
indicates the fragmented nature of the corresponding ground states. Another
notable feature in figure 2 is that the overall profile of $1/I_C$ exhibits
a valley around $L/N=1.0,1.8,$ and $2.4$. This is consistent with the broad
peaks of $n_j$ at $L\approx 40,70$ and $90$ for $j=1,2$ and $3$,
respectively, as shown in figure 1. The number fluctuations $\Delta n_j$ are
in the order of $O(n_j^{1/2})$ for these peaks, exhibiting a {\em local}
characteristic of a single coherent condensate.

{\it The intrinsic spontaneous symmetry breaking of a fragmented state}%
.---Let us now consider the stability of such a fragmented ground state. In
Ref. \cite{rokhsar98,ho}, the authors argued that the fragmented state is
inherently unstable to the formation of a single condensate of well-defined
phase. The essential idea is that even a weak perturbation that breaks the
conservation laws will rapidly generate phase coherence, modifying the
density matrix deterministically to give a unique macroscopically occupied
SP state. To support this point, we first show that the fragmented state and
its corresponding single condensate state have the same energies in the
limit of $N\rightarrow \infty $, up to the order of $O(gN).$ A similar
conclusion has been reported by Jackson {\it et al. }for the special case of 
$L=2N$ \cite{mft-vs-ex}. In figure 3, the interaction energy $V_{int}$ in
units of $gN^2$ is plotted as a function of $L/N$ for a system of $N=20,40,$
and $60$ bosons. As $N$ increases, $V_{int}$ becomes closer in value to that
of a single condensate $V_{int}^{mf}$, as predicted by the MFT\cite{mft}.
The inset shows the energy difference $\Delta V_{int}=V_{int}^{mf}-V_{int}$
in units of $gN$. It is readily seen that all the $\Delta V_{int}$ with
different $N$ are approximately located on a {\em universal} curve. This
strongly suggests that $\Delta V_{int}$ can be described by an approximate
form 
\begin{equation}
\Delta V_{int}=\alpha gN\ll \hbar \omega ,
\end{equation}
where in the thermodynamic limit the factor $\alpha \sim 1$ depends on $L/N$
only and the inequality comes from our assumption of weak interaction. As a
result, even a perturbation of order $O(1/N)$ can be enough to drive the
fragmented state into a single condensate state. This fact clearly indicates
that the fragmented state will spontaneously break whatever the
fragmentation was permitted by cylindrical symmetry in the first place.

This result can be understood in another way by considering the conditional
probability distributions (CPDs) \cite{corr} that give the density
correlation among bosons. We define the CPD for finding one boson at ${\bf r}
$ given another at ${\bf v}_0$ as 
\begin{equation}
{\cal P}({\bf r\mid v}_0)=\frac{\left\langle \Psi _{GS}\right|
\sum\nolimits_{i\neq j}\delta ({\bf r}-{\bf r}_i)\delta ({\bf v}_0-{\bf r}%
_j)\left| \Psi _{GS}\right\rangle }{(N-1)\left\langle \Psi _{GS}\right|
\sum\nolimits_j\delta ({\bf v}_0-{\bf r}_j)\left| \Psi _{GS}\right\rangle }.
\label{cpd}
\end{equation}
Unlike the usual density distribution that is cylindrically symmetric under
rotational invariant confinement, the CPD is asymmetric and reflects an {\em %
intrinsic} density distribution \cite{cpd-note}.

What will an inherently unstable fragmented state evolve into if a weak
perturbation is switched on? One may expect that the system will rapidly
change into a state having the same intrinsic density distribution as the
fragmented state, and simultaneously generate phase coherence \cite{ho}. In
view of this, the CPD gives the tendency of a system's evolution and can be
regarded as a measurement of the possible spontaneous symmetry breaking.

In figure 4, we show the $L$ dependence of the CPDs for a system of $N=40$
bosons. As expected, we observe the successive vortex-like patterns of
differing symmetries, which are in good qualitative agreement with the
mean-field calculations \cite{mft}. Both of them show a gradual transition
for the formation of one- (fig.4a) and two-vortex-like (fig.4b) states in
contrast to the rapid appearance of the three-vortex-like state (fig. 4c) 
\cite{mft}. As mentioned above, we identify this similarity as a signal of
spontaneous symmetry breaking of fragmented states.

On the other hand, one should not confuse CPDs with the ``true'' vortex
patterns predicted by the MFT \cite{mft}. The latter has phase coherence,
which is not just well-defined in CPDs. Besides this, they have a different
physical mechanism for the vortex emergence with the increasing $L$. For
example, our results seem to show that the one- and two-vortex are produced
at the center of the cloud of condensate, in apparent contradiction to the
prediction of the MFT that the vortex enters the cloud from the low-density
periphery. These differences may be resolved through the Josephson tunneling
experiment suggested by Leggett and Sols \cite{leggett}. Certainly, more
accurate theoretical studies on the fragmented state are required.

In conclusion, we have studied the ground state of a weakly interacting $N$%
-Boson system with a given angular momentum. We propose that the ground
state is generally a fragmented condensate state, which is rather fragile in
response to a weak asymmetric perturbation. By calculating the corresponding
CPDs, we manifest this intrinsic instability. A comparison with the
mean-field results is also given.

We thank Y. Zhou, Y.-X. Miao, and C.-P. Sun for their stimulating
discussions. X.-J. Liu was supported by the NSF-China.

\begin{center}
{\bf Figures Captions}
\end{center}

Fig. 1. (color) Values of $n_j$ (in blue) and $\Delta n_j$ (in red) of five
lowest SP oscillator states as a function of $L$ for a system of $N=40$
bosons with $j_{\max }=6$. The cases of $j=5$ and $6$ are not shown due to
their low occupancy.\newline

Fig. 2. $1/I_C$ versus $L/N$ for a system of $N=30,40,50,$ and $60$ bosons.
Small oscillations at $L/N>1.6$ are caused by the finite $N$ effect. The
overall profile of $1/I_C$ is nearly independent of $N$, thus $1/I_C$ can be
used to qualitatively measure the number of macroscopic occupation numbers
in the limit of $N\rightarrow \infty $. \newline

Fig. 3. (color) $V_{int}$ in units of $gN^2$ as a function of $L/N$ for a
system of $N=20$ (in green)$,40$ (in blue)$,$ and $60$ (in red) bosons. For
comparison, the $V_{int}$ predicted by MFT is also depicted by the dark
solid line. The inset shows the energy difference $\Delta
V_{int}=V_{int}^{mf}-V_{int}$ in units of $gN.$ Note that all the $\Delta
V_{int}$ with different $N$ are approximately located on a universal curve.%
\newline

Fig. 4. Selected CPDs for a system of $N=40$ bosons. (a), (b), (c) and (d)
correspond to the emergence of vortex-like patterns with $p$-fold symmetry ($%
p=1,2,3,4$). In each panel, $L$ increases in steps of one unit, and the
starting value of $L$ in (a), (b), (c), and (d) is $33,62,79,$ and $108$,
respectively. The values of $x$ and $y$ in each subplot range from $-3.0$ to 
$+3.0$. The given point ${\bf v}_0$ is $(0$, $1.0).$ For large $N$, the CPD
is nearly independent of ${\bf v}_0$.


\begin{references}
\bibitem{jmaxnote}  In the mean-field calculation, the occupancy of SP
states with $j>6$ is very low for $L\leqslant 2N$ \cite{mft}. For $%
2N<L\leqslant 3N$, our numerical results with $N=20$ show that $n_j/N<0.03$ (%
$j>6$), therefore including such states does not alter our results
qualitatively.
\end{references}
\end{document}